\documentclass{article}

\usepackage{icrctc07}

\def\simleq{\; \raise0.3ex\hbox{$<$\kern-0.75em \raise-1.1ex\hbox{$\sim$}}\; }
\def\simgeq{\; \raise0.3ex\hbox{$>$\kern-0.75em \raise-1.1ex\hbox{$\sim$}}\; }

\newcommand{\GeV}{{\rm GeV}}
\newcommand{\TeV}{{\rm TeV}}

\newcommand{\kpc}{{\rm kpc}}
\newcommand{\pc}{{\rm pc}}
\newcommand{\cm}{{\rm cm}}
\newcommand{\km}{{\rm km}}
\newcommand{\muG}{\mu{\rm G}}

\newcommand{\s}{{\rm s}}

\newcommand{\sr}{{\rm sr}}

\title{Gamma-ray and neutrino diffuse emissions of the Galaxy above the TeV}

\authors{Carmelo Evoli$^1$, Dario Grasso$^2$, Luca Maccione$^{1,3}$}

\afiliations{$^1$ SISSA, via Beirut, 2-4, I-34014 Trieste\\
$^2$ INFN, Sezione di Pisa, Largo Bruno Pontecorvo, 3, I-56127 Pisa\\
$^3$ INFN, Sezione di Trieste, Via Valerio, 2, I-34127 Trieste}
\email{evoli@sissa.it, dario.grasso@pi.infn.it, maccione@sissa.it}
\abstract{In this contribution we will  discuss recent results concerning the intensity and the angular distribution of the gamma-ray and neutrino emissions  as should be originated from the hadronic scattering of cosmic rays (CR) with the interstellar medium (ISM). We assumed that CR sources are supernova remnants (SNR) and estimated the spatial distribution of primary nuclei by solving numerically the diffusion equation.  For the ISM, we considered recent models for the 3D spatial distributions of molecular hydrogen. Respect to previous results, we find the secondary gamma-ray and neutrino emissions to be more peaked along the galactic equator and in the galactic centre which improves significantly the perspectives of a positive detection. We compare our predictions with the experimental limits/observations by MILAGRO and TIBET (for the gamma-rays) and by AMANDA-II (for the neutrinos) and discuss the detection perspectives for a km3 neutrino telescope to be built in the North hemisphere.}

\begin{document}
\maketitle

\section{Introduction}

Several orbital observatories  (see \cite{Bloemen:89} for a review), especially EGRET \cite{Hunter:97,Cillis:05},  found  that, at least up to $10~\GeV$,  the Galaxy is pervaded by a  $\gamma$-ray diffuse radiation.  
While a minor component of that emission is likely to be originated by  unresolved point-like sources,
the dominant  contribution is expected to come from the interaction of  galactic cosmic rays (CR) with 
the interstellar medium (ISM).  Since the spectrum of galactic  CR extends up to the EeV,  the spectrum of 
$\gamma$-ray diffuse galactic emission should  continue well above the energy range probed
by EGRET.  That  will be soon probed by GLAST \cite{Glast} up to $~300~\GeV$ and by air shower arrays 
(ASA) (e.g.  MILAGRO \cite{Milagro:obs,Abdo:2006} and TIBET \cite{Tibet:obs})  above the TeV.   

Above the GeV,  the main $\gamma$-ray emission processes are expected to be the decay of $\pi^0$ produced by 
the scattering of CR nuclei onto the diffuse gas  (hadronic emission)   and the Inverse Compton (IC) emission of relativistic electron colliding onto the interstellar radiation field (leptonic emission).   
It is  unknown, however,  what are the relative contributions of those two processes  and how they  change with  the energy and  the position in the sky (this is so called {\it hadronic-leptonic degenerary}). 
Several numerical simulations have been performed in order  to interpret EGRET as well as 
forthcoming measurements at high energy (see e.g. \cite{Berezinsky:93,Aharonian:00,Strong:04} ). Generally,  those simulations predict  the hadronic emission to be dominating between 0.1 GeV and few TeV, 
while  between 1 and 100  TeV a comparable, or even larger IC contribution may be allowed.    

The 1-100 TeV energy range, on which we focus here,  is also interesting from the point of view of neutrino astrophysics. In that energy window neutrino telescopes (NTs)  can look for  up-going muon
neutrino and reconstruct their arrival direction with an angular resolution better than $1^\circ$.  
 Since hadronic  scattering give rise to $\gamma$-rays and neutrinos in a known ratio, the  possible measurement  of the  neutrino emission from the Galactic Plane (GP) may allow to get rid of the hadronic-leptonic  degeneracy.  
 
 In this contribution we discuss the main  results of  a recent work were we modelled  the 
$\gamma$-ray and neutrino diffuse emission of the Galaxy due to hadronic scattering \cite{our_paper}. 
Our work  improves previous analysis  under several aspects which  concern 
 the distribution of CR sources;    the way we treated CR  diffusion by  accounting for spatial variations of the diffusion coefficients;    the distribution of the atomic and molecular hydrogen.

\section{The spatial structure of the ISM} \label{sec:galaxy}

In order to assess the problem of the propagation of CRs and their interaction with the  ISM we need the knowledge of three basic physical inputs, namely:
the distribution of SuperNova Remnants (SNR) which we assume to trace that of  CR sources; the properties of the  Galactic Magnetic Field (GMF) in which the propagation occurs; the distribution of the diffuse gas providing the target for the production of $\gamma$-rays and neutrinos through hadronic interactions.  In the following we assume cylindrical symmetry  and adopt the Sun galactocentric distance $r_\odot = 8.5~\kpc$. 

\subsection{\it The SNR distribution in Galaxy}\label{subsec:snr}

Several methods to determined the  SNR  distribution in the Galaxy  are discussed in the literature (see e.g. that based on the surface brightness - distance relation  \cite{Case:1998qg}). 
Here we adopt a SNR distribution a distribution as inferred  from observations of related objects, such as pulsars or progenitor stars, as done e.g. in \cite{Ferriere} which is 
less plagued from sistematics  and agrees with that inferred from the distribution of
radioactive nuclides like of $^{26}$Al . 
A similar  approach was followed in \cite{Strong:04b}  where, however, the contribution of type I-a SNR (which is dominating in the inner 1 kpc) was disregarded.  

\subsection{\it Regular and random magnetic fields}\label{subsec:mf}

The Milky Way, as well as other spiral galaxies, is known to be permeated by large-scale,
so called {\it regular}, magnetic fields as well as  by a random, or turbulent,  component. 
The orientation and strength of the regular field is measured mainly by means of Faraday Rotation Measurements (RMs) of polarised radio sources.  From those observations it is known that  the  regular field in the disk of the Galaxy  is prevalently oriented along the GP.  Following \cite{Han:94,Han:06}
we adopt the following analytical distribution for the disk and  the halo: 
\begin{equation}
\label{Breg}
B_{\rm reg}(r,z) = B_0 \exp \left\{ - \frac{r - r_\odot}{r_B}\right\} \; 
\frac{1}{2\cosh(z/z_r)}~,
\end{equation}
where  $B_o \equiv B^{\rm disk}_{\rm reg}(r_\odot, 0)  \simeq 2~\muG$ is  the  strength at the Sun circle.
 The parameters $r_B$ and $z_r$ are poorly known. However we found that our final results are practically independent on their choice.  In the following we adopted $r_B = 8.5 ~\kpc$ and $z_r  = 1.5~\kpc$. 
 
More uncertain are the properties of  the turbulent  component of the GMF.  
 Here we assume that it strength follows the behaviour  
\begin{equation}
\label{Bran}
B_{\rm ran}(r,z) = \sigma(r) \;B_{\rm reg}(r,0) \;\frac{1}{2\cosh(z/z_t)}\; .
\end{equation}
where  $\sigma(r)$ parematrise the turbulence strength.  Here we assume $z_t = 3~\kpc$.
From polarimetric measurements and RMs is known GMF are chaotic on all scales below $L_{\rm max} \sim 100~\pc$.
The power spectrum of the those fluctuations is also poorly known.  In \cite{our_paper} we considered both a Kolmogorov ($B^2(k) \propto k^{-5/3}$)  and a Kraichnan ($B^2(k) \propto k^{-3/2}$) power spectra.

\subsection{\it The gas distribution}\label{subsec:gas}

The model which consider here is based on a suitable combination  of different analyses which have been separately performed for the disk and the galactic bulge. For the ${\rm H}_2$ and HI distributions in the bulge we use a detailed 3D model recently developed by Ferriere et al. \cite{Ferriere:07} on the basis of several observations. For the molecular hydrogen in the disk we use the well known Bronfman's et al. model \cite{Bronfman:88}. For the HI distribution in the disk, we adopt Wolfire et al. \cite{Wolfire:03} 2-dimensional model. 
In the following we will assume that  helium is distributed in the same way of  hydrogen nuclei.

\section{CR diffusion} \label{sec:diffusion}

The ISM is a turbulent magneto-hydro-dynamic (MHD) environment. Since the Larmor radius of high energy nuclei is smaller than $L_{\rm max}$, the propagation of those particles takes place in the spatial diffusion regime. 
The diffusion equation describing such a propagation is (see e.g. \cite{Ptuskin:93}) 
\begin{equation}\label{j}
- \nabla_i\left( D_{ij} (r,z) \nabla_j N(E,r,z)
\right)  = Q (E,r,z) 
\end{equation}
where $N(E,r,z)$ is the differential CR density averaged over a scale larger than $L_{\rm max}$, $Q (E,r,z) $ is the CR source term  and $D_{ij} (E,r,z)$ are the spatial components of the diffusion tensor. 
In  the energy range considered in our work  energy loss/gain can be safely neglected.
Since we assume cylindrical symmetry the only physically relevant components of the diffusion coefficients are  $D_{\perp}$ and $D_A$, respectively the diffusion coefficient in the direction perpendicular to ${\bf B}_{\rm reg}$ and the antisymmetric (Hall) coefficient.  We adopted  expressions for those coefficients as derived by Montecarlo simulation of charged particle propagation in turbulent magnetic fields \cite{Candia:04}.
Respect to other works, where only isotropic diffusion was considered and a 
a mean value of the diffusion coefficient was estimated  from the observed 
secondary/primary ratio of CR nuclear species (see e.g. \cite{Strong:04}),  our approach 
offers the advantage to provide the diffusion coefficients {\it point-by-point} at any energy.   
We solved the diffusion equation using the Crank-Nicholson method  by imposing $N(E) = 0$ at the edge of the MF turbulent halo and by requiring that it matches the observed CR spectra at the Earth position for most abundant nuclear  species.
   
\section{Mapping the $\gamma$-ray and $\nu$ emission}\label{sec:neutrinos}
\begin{figure}[!ht]
 \begin{center}
\noindent
 \includegraphics[scale=0.45]{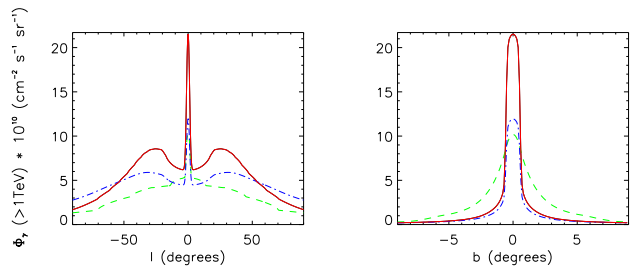}
\end{center}
 \caption{The $\gamma$-ray flux profiles along the GP (left panel) and along $l = 0$ (right panel)
 for $E > 1~\TeV$, averaged over $1^\circ \times 1^\circ$ angular bins.  The continuos (red), dot-dashed (blue) and  dashed (green) curves correspond respectively to: our work, a model with the same gas density distribution  but a uniform CR flux, taken to match  direct observations at the Earth;  the model considered in  \cite{Berezinsky:93}. 
 The corresponding neutrino flux can be obtained by dividing this diagram by 3.1.}
 \label{fig:nuprof_comp}
\end{figure}

Under the assumption that the primary CR spectrum is a power-law and that the differential cross-section follows a scaling behaviour (which is well justified at the energies considered in this  paper), the $\gamma$-ray (muon neutrino) emissivity due to hadronic scattering can be written as 
\begin{eqnarray}
 \frac{dn_{\gamma~(\nu)} (E;~ b,l)}{dE} 
  &\simeq  & f_N ~\sigma_{pp}  \; Y_{\gamma}(\alpha)\\
  &&\int  {\rm d}s  \; I_p(E_p;~ r,z)~n_H(r,z) 
  \nonumber
  \end{eqnarray}
 Here  $I_p(E_p;~ r,z)$ is the proton CR differential flux at the position $r, z$  as determined by 
 solving the diffusion equation;  $\sigma_{pp}$ is the 
 $pp$ cross-section;  $Y_ {\gamma}  \simeq 0.04$ and $Y_ {\nu_{\mu} + {\bar \nu}_\mu}  \simeq 0.01$  
 are the $\gamma$-ray  and muon neutrino yields  respectively as obtained for  a proton spectral slope $\alpha = 2.7$ (see Fig. 7 in \cite{our_paper,us:05} and ref.s therein);  the factor $f_N \simeq 1.4 $ represents the contribution from all other nuclear species both in the CR and the ISM;  $s$ is the distance from the Earth; 
 $b$ and $l$ are the galactic latitude and longitude.  
Here we assume that the slope of the CR injection spectrum is 2.2 and that the GMF turbulent spectrum 
has a Kraichnan spectrum. 
Fluxes obtained by using different spatial structure and spectra of the turbulent magnetic field differ 
by a factor 2 at most.
In \ref{fig:nuprof_comp} we compare our results with those obtained in \cite{Berezinsky:93}.

\section{Discussion}

First of all, we  compare our results  with EGRET observations in the $4- 10~\GeV$ energy range \cite{Cillis:05}. 
That is possible  since, already for  $< 10~\TeV$,  nuclei propagation takes place deep into the spatial diffusion regime so that the behaviour  of the diffusion coefficients do not change going to lower energies and our results can be safely extrapolated. As we showed in  \cite{our_paper}, 
our predictions are in good agreement with EGRET measurements along the GP, but a small
deficit which can be  easily explained in terms of IC.  
Then we can reliably compare our results with measurements performed above the TeV
with air shower arrays (see Tab. 1)  and NTs. 
 We found that  with the exception of Cygnus (where one or more  sources are likely to increase the 
 local CR density) in all other regions  we predict  fluxes which are significantly below the experimental limits. 

Concerning neutrinos,  the only available upper limit on the neutrino flux  from the Galaxy has been obtained by the AMANDA-II experiment  \cite{Kelley:05}.  Being located at the South Pole,
AMANDA cannot  probe the  emission from the GC.  In  the region  
$33^\circ < l < 213^\circ, \ |b| < 2^\circ$, and assuming a spectral index $\alpha = 2.7$, 
 their present constraint is  $\Phi_{\nu_{\mu} + {\bar \nu}_\mu} (> 1~\TeV) < 3.1 \times 10^{-9} 
~(\cm^2~ \s~ \sr)^{-1}$. According to  our model the expected flux  in the same region is   
$\Phi_{\nu_{\mu} + {\bar \nu}_\mu} (> 1~\TeV) \simeq  4.2 \times 10^{-11} 
~(\cm^2~ \s~ \sr)^{-1}$.  That will be hardly detectable even by IceCube. 
Slightly more promising are the perspectives of a $\km^3$ neutrino telescope to be built in the North 
hemisphere.  In \cite{our_paper} we found, however, that the Km3Net project may be able to detect an excess in direction of the  GC only if a significant CR over-density  is present 
in that region.

\begin{table}[!b]
\caption{\label{tab:exp_comp} In this table our predictions for the mean $\gamma$-ray flux in some selected regions of the sky are compared with some available measurements. Since measurements' errors are much smaller than theoretical uncertainties they are not reported here.}
\label{tab:comp}
\centering
\begin{tabular}{@{}|cccc|}
\hline
sky window  & $E_{\gamma}$ & \multicolumn{2}{c}{$\Phi_\gamma(>E_\gamma)~(\cm^2~\s~ \sr)^{-1}$}\\
\hline
            &             &   our model & measurements \\
\hline
$|l|  < 10^\circ,\ |b| \le 2^\circ$ & $4~\GeV$ & $ \simeq  4.7 \times 10^{-6}$ &$\simeq  6.5\times 10^{-6}$ \cite{Cillis:05}\\
$20^\circ \le  l \le 55^\circ,\ |b| \le 2^\circ$ & $3~\TeV$ & $\simeq 5.7 \times 10^{-11}$ & $\le 3 \times 10^{-10}$  \cite{Tibet:obs} \\
 & $4~\GeV$ & $\simeq  4.4 \times 10^{-6}$ & $\simeq  5.3 \times 10^{-6}$  \cite{Cillis:05}\\
$73.5 ^\circ \le  l \le 76.5^\circ,\ |b| \le 1.5^\circ$ &  $12~\TeV$ & $\simeq   2.9\times 10^{- 12}$ & $\simeq  6.0\times 10^{-11}$  \cite{Abdo:2006}\\
 &$4~\GeV$ & $\simeq  2.4 \times 10^{-6}$ & $\simeq  3.96 \times 10^{-6}$ \cite{Cillis:05} \\
$140^\circ < l < 200^\circ, \ |b| < 5^\circ$ & $3.5~\TeV$ & $\simeq  5.9 \times 10^{-12}$ & $\le 4 \times 10^{-11}$  \cite{Milagro:obs}\\
 & $4~\GeV$ & $\simeq  5.9 \times 10^{-7}$ & $\simeq 1.2\times 10^{-6} $  \cite{Cillis:05} \\
\hline
\end{tabular}
\end{table}
 
\bibliography{icrc0348}
\bibliographystyle{plain}

\end{document}